\RequirePackage{fix-cm}
\documentclass{svjour3}                     
\smartqed  
\usepackage{graphicx}
\usepackage{url}
\usepackage{natbib}
\usepackage{color}
\usepackage{multirow}
\newcommand{\new}[1]{{\color{black}#1}}
\newcommand{\newnew}[1]{{\color{black}#1}}
\begin{document}

\title{\textit{Judging the algorithm}
}
\subtitle{\new{Algorithmic accountability on} the risk assessment tool for intimate partner violence in the Basque country}

\titlerunning{Judging the algorithm}        

\author{Ana Valdivia \and Cari Hyde-Vaamonde \and Julián García Marcos}
\authorrunning{Valdivia et al.} 

\institute{Ana Valdivia \at University of Oxford, corresponding author: \email{ana.valdivia@oii.ox.ac.uk} \and Cari Hyde-Vaamonde \at King's College London \and Julián García Marcos \at Euskal Herriko Unibertsitatea}


\maketitle

\begin{abstract}
This \new{paper discusses \newnew{an algorithmic} tool introduced in the Basque Country (Spain) to assess the risk of \newnew{intimate partner violence}. \newnew{The algorithm was introduced to overcome the lack of human experts by calculating automatically the level of violence based on psychometric features such as controlling or violent behaviour.}

Although  critical literature on risk assessment tools for domestic violence mainly focuses on English-speaking countries, this paper offers an algorithmic accountability analysis of \newnew{its} risks, harms, and limitations \newnew{in this Spanish region}. We propose a \newnew{trans}disciplinary approach from a statistical and legal perspective. This approach unveils issues and limitations that could lead to unexpected consequences for individuals suffering from \newnew{partner} violence. Moreover, while the tool has a high error rate on severe cases, i.e. cases where the aggressor could murder his partner (5 out of 10 high-risk cases are misclassified as low risk), there is a lack of appropriate legal guidelines for judges, the end-users of this tool. The paper concludes that \newnew{this tool needs to be urgently} assessed by independent and transdisciplinary experts to better mitigate algorithmic \newnew{harms}.}
\end{abstract}


\section{Introduction}
\label{intro}
In 2018, \new{María}\footnote{\new{Not her real name.}} reported her ex-husband to the police authorities in the Basque country (Spain). She suffered from domestic violence perpetuated by her ex-husband.\footnote{According to United Nations (UN), the definition of domestic violence is: \textit{`Domestic abuse, also called ``domestic violence'' or ``intimate partner violence'', can be defined as a pattern of behavior in any relationship that is used to gain or maintain power and control over an intimate partner. It is considered a type of gender-based violence''}. (See: \url{https://www.un.org/en/coronavirus/what-is-domestic-abuse} (Last accessed January, 16 2023)).} During the report, the police authorities asked \new{María} several questions such as: \textit{Are you an immigrant?} \textit{Is the aggressor showing very intense jealousy or controlling behaviours to you?} \textit{Is your life in danger?} The output of these questions fed to an algorithmic tool that automatically quantified the severity of domestic violence (see Appendices~\ref{app:a} and~\ref{app:b} \new{to observe all the questions of this risk assessment tool}). According to the authors of this algorithmic tool: `it empirically establish[es] risk markers of severe injuries and homicide in intimate partner violence'~\citep{echeburua2009assessing}. In the case of \new{María}, the algorithmic output assessed that her ex-husband had a low-risk of severe domestic violence. When the report of \new{María}'s case containing both the context and the algorithmic output, arrived into the courtroom, the judge observed that this case was rather a high-risk case. Although there were no death threats and \new{María} did not perceive risk of death when she was interrogated by the police, the perpetrator was stalking her and her daughter, which negatively impacted on their mental health. This factor was not considered nor assessed by the algorithmic tool. In this case, the perpetrator did \new{not} show any explicit violent behaviour which would lead this tool to automatically assess the risk of severe violence as high given its design. Rather he stalked them by walking around their house or buzzing the entryphone of their apartment---which \new{was} not considered by the \new{algorithmic tool}.

Algorithms have been introduced in police stations and courtrooms to facilitate the decision-making process of law enforcement actors. There is an increasing number of algorithmic-based solutions adopted by law enforcement authorities to enhance efficiency and effectiveness~\citep{interpol2020ai, singh2018handbook}. Yet most of these data-driven solutions are permeating judicial systems in a quiet and understated way. In some cases, end-users \new{such as} judges and police officers are using algorithmic outputs as evidence without having a deep understanding about its risks and limitations. Moreover, these algorithms are often \new{not evaluated} once they are implemented \new{in real scenarios}. Although in most cases algorithmic tools are implemented to improve efficiency \new{and bring potential benefits}, \new{there are risks associated with these tools. For instance, algorithmic tools can perpetuate inefficiency due to pitfalls in the design or implementation}~\citep{valdivia2022paradox}. In some cases, the lack of information, training and technical knowledge by \new{law enforcement agents} on algorithmic systems can \new{exacerbate the risks and limitations that arise when using these tools}. \new{As the legal and computer science scholar~\citeauthor{binns2022human} eloquently states: `it is not obvious how these different elements of justice, and the humans and machines which are supposed to serve them, can be combined without undermining each other' (2022, p. 208).} \new{Indeed, the collaboration between humans and algorithms cannot be \new{seamless in each case} 
and the literature advocates for a need in better scrutinising the benefit-risk trade-off of algorithmic governance in law enforcement authorities in order to mitigate harms~\citep{kleinberg2018discrimination, ludwig2021fragile, de2020case}}. 

While part of the \new{critical data literature on risk assessment tools for intimate partner violence} has focused on analysing algorithmic systems within English-speaking countries, \new{other geographies remain unexplored. To the best of our knowledge, this paper presents the first critical analysis of the tool used to assess partner violence in a region of Spain.} Judges \new{in the Basque country, an autonomous community in northern Spain,} have been exposed to an algorithmic tool that assesses the risk of intimate partner violence, such as the \new{María}'s case presented before. Without sufficient training or knowledge on the design, implementation and evaluation of algorithms, this tool was proposed to overcome the lack of police experts on domestic violence. The 
\textit{Escala de Predicción de riesgo de Violencia grave contra la pareja} (Intimate Partner Femicide and Severe Violence Assessment) tool (EPV hereinafter) consists of 20 items that discriminate between severe and non-severe domestic violence. It was primarily designed to assist police officers on the assessment of the risk of suspects to inflicting severe injuries and femicide.\footnote{Domestic, or intimate partner violence occurs irrespective of the gender of the individuals involved. Not all aggressors are male and not all victims are female. In this article we analyse the impact of EPV which focus specifically on intimate partner violence, but this is an unhelpful reduction of the overall picture.} 

This paper presents \new{an algorithmic accountability} framework to critically analyse the EPV from a socio-technical and legal perspective. Through the lens of a team \new{made by} a critical computer scientist, legal scholar and a judge in the Basque court system, we investigate risks, harms and \new{limitations} \new{arising from this} algorithmic tool. From a socio-technical perspective, the EPV is a classic weighting system \new{tool} that assesses the risk of violence. \new{However, we identify serious risks associated with the design of this algorithm, such as its performance on high-risk cases.} Our framework unveils that more than 50\% high risk cases were classified as low-risk. \new{On the other hand, from a legal perspective,} the consideration of the results of the algorithm without judicial training or clear guidelines on its evidential basis appears at odds with the judge's role, in law, to assess risk themselves. This algorithm is not a report of objective data, but instead synthesises an amalgam of human, subjective, judgements, which are not open to scrutiny in the way that witness evidence is subject to scrutiny. This is especially concerning for the individuals involved, but also risks broader consequences for the the efficacy of the justice system itself, which to a great extent relies on trust and cooperation. Through discussion with \new{the judge} calls into question the benefits that this tool is bringing to the Basque courtrooms. While judges had little knowledge about the design process of the EPV, they are exposed to the algorithmic output in their daily lives which has an impact on the court decision. \new{Judicial procedures impacted by \textit{techno-solutionism}}\footnote{A defined by \newnew{\citet{morozov2013save}}, tecno-solutionism refers to `recasting all complex social situations either as neatly defined problems with definite, computable solutions or as transparent and self-evident processes that can be easily optimized'~\citep[p. 5]{morozov2013save}.} are in danger of being influenced by \new{algorithmic assessments} and \new{could undervalue} their experience as \new{legal experts} in domestic violence.



The structure of this paper is as follows: in Section~\ref{sec:back} we conduct a literature review on algorithmic assessment tools that evaluate the risk of violence since the 1980s. In Section~\ref{sec:howepv} we provide a detailed description about how this algorithm was designed and introduce in the Basque courtrooms. In Section~\ref{sec:tech} we analyse the EPV tool from a statistical and technical perspective. We offer a legal analysis in Section~\ref{sec:legal}. The paper moves on by identifying its risks, limits and harms from a socio-technical and legal perspective in Section~\ref{sec:lim}. In Section~\ref{sec:discussion} we offer a discussion about this tool moving the critical debate beyond bias and algorithmic opacity. We claim that this tool has several design and implementation errors that could potentially impact on individuals that are suffering from intimate partner violence in the Basque country. We conclude that this tool needs to be urgently audited in Spain to avoid and mitigate further algorithmic harms.

\section{\new{Risk assessments for quantifying violence: statistical-based tools outperforming human judgments?}}
\label{sec:back}


The use of risk assessment tools to \new{understand} violence \new{from a quantitative perspective} has a long-standing history. Statistical predictions began in the 1980s through the analysis of risk factors by `predicting an individual’s behavior on the basis of how others have acted in similar situations or on an individual’s similarity to members of violent groups'~\cite[p. 26]{campbell1995assessing}\newnew{---}see also~\cite{miller1988predictions}\newnew{---}. \new{Scholars} from different disciplines involving psychology, criminology and statistics have analysed how violence could be quantified and even predicted. \newnew{For instance,}~\citeauthor{gottfredson1988violence} proposed several statistical strategies such as bootstrapping\footnote{\new{Statistical technique that resamples an original dataset to create simulated ones. The resampled datasets are the same size as the original dataset and only contain values that exist in the original set.}} and contingency tables\footnote{\new{Table that compares two features. The values are frequencies for each unique combination of the two compared variables.}} to overtake human judgments and improve the justice and mental health system, performing `more efficiently and more effectively'~\newnew{(1988, p. 318)}. These authors contended that violence can and should be predicted and justified the use of statistical tools by acknowledging that scholars such as~\cite{takeuchi1981behavioral} advised that human judgements are highly fallible. \new{In the specific case of domestic-based violence, \cite{campbell1995assessing} designed the first risk assessment tool \new{almost thirty} years ago. \new{Allegedly, the main purpose of these tools is to `reduce harm to female victims of intimate partner violence and their children' and `facilitate the gathering of detailed and relevant information about the victim and the perpetrator in intimate partner violence cases'~\cite[p.~19]{risk2019EU}}. This danger assessment instrument was proposed to be used in clinical contexts, primary care and civil justice settings~\citep{campbell1995assessing}}.\footnote{See: \url{https://www.dangerassessment.org/} (Last accessed, February 16, 2022).} \new{It was composed of 20 features associated with the risk of murdering a partner}. Every item \new{was} evaluated with a binary output (yes/no) which \new{was} in turn associated \new{with} a \new{binary score} (1/0). The final score, which \new{was} the sum of each response for each \new{feature}, \new{assessed} the \new{risk} of intimate partner homicide. In the last decades, these tools have evolved and become more sophisticated due to the development of machine learning techniques and \new{the ability to gather, store, and process} more data. \new{For instance,}~\cite{rodriguez2020modeling} compared different machine learning classifiers (logistic regressions, random forest, support vector machines and Gaussian process \new{algorithms}) to estimate the number of domestic violence complaints. Using a \new{classic data analysis} approach, \newnew{\citet{cumbicus2021data}} proposed data mining techniques to better understand \new{the} underlying causes of intimate partner violence.

\new{The critical literature on risk assessment tools has claimed that `we have limited understanding of their properties: most notably, whether and how they actually improve decision-making'~\cite[p. 97]{green2019disparate}. From a US perspective, scholars~\citeauthor{thomas2022automating} found that algorithmic risk assessments in pretrial adjudications are unconstitutional: `their opacity, biases, judicial influence, and racially disparate treatment of Black and Latino defendants, all of whom are legally innocent, likely do not pass muster under the Equal Protection framework' \newnew{(2022, p. 407).} In their study, they advocated for the abolition of these tools due to potential violations of fundamental rights. Moreover, they claimed that not even the mitigation of bias would render them legal tools in the US. Another interesting study analysing the risks of assessment tools in court decision was published by\newnew{~\citet{albright2019if}}. \newnew{This scholar} analysed a predictive tool implemented for bail decisions in Kentucky (US) and its interaction with judges, concluding that judges made different decisions influenced by the race of the subject, even when the algorithmic score was the same~\citep{albright2019if}. More recently,~\cite{stevenson2023counterintuitive} examined the results subsequent to implementing a risk assessment tool to evaluate sexual recidivism in Virginia (USA). \newnew{Their findings highlighted that rape sentences decrease because `the risk assessment served as a sort of second opinion that made judges feel more comfortable granting leniency for those in the lowest risk category'~\citep[p.~70]{stevenson2023counterintuitive}. If judges are mistaken with their decisions because a person goes to re offend, they can point out to the faulty score and mitigate internal guilt~\citep[p.~71]{stevenson2023counterintuitive}.}}

\new{Despite these key studies that evaluated the performance of risk assessment tools in US courts, there is a general lack of algorithmic accountability from a statistical perspective on risk assessment tools for intimate partner violence. In fact, this evidence has been identified by \new{the}~\citeauthor{risk2019EU} claiming in the report \textit{Risk assessment and management of intimate partner violence in the EU} that: `[t]here is a relatively small body of empirical evidence to evaluate tools that assess the risk of intimate partner violence'~\newnew{(2019, p. 32)}. This question is highly relevant, as there is evidence suggesting that risk assessment tools do not always outperform human judgments~\citep{dressel2018accuracy}---contrary to what scholars developing risk assessment tools claimed\newnew{---}see ~\citep{takeuchi1981behavioral} or ~\citep{gottfredson1988violence}\newnew{---}. In fact, \newnew{\citet{nicholls2013risk}} \new{have identified that these tools have weak to moderate predictive accuracy.} It is widely known that the use of algorithmic tools involves risks, harms and limitations that need to be taken into account during its design and implementation in real life. For instance, some of these limitations are related with the demystification of the neutrality and objectivity of algorithmic and statistical tools~\citep{birhane2021values}, the unintended discrimination and disparate impact~\citep{chouldechova_2017_fair, angwin2016machine} and the influence that automated tools have on human decisions~\citep{green2019disparate, green2021algorithmic, binns2022human}}. 


\new{While risk assessment tools for intimate partner violence are widespread across the globe---for instance, see~\citep{cumbicus2021data} for Ecuador,~\citep{amusa2020predicting} for South Africa, and~\citep{dehingia2022help} for India)}, most of these studies are adopting tools designed for the US or Canadian context. Europe is not an exception~\citep{risk2019EU}. This fact could negatively impact on the tool's performance \new{given that intimate partner violence behaviour strongly depends on the context~\citep{zark2022cross}}---which contributes to the limitations of risk assessment tools discussed before. \new{In the specific case of Spain}, \new{the Ministry of Interior adopted the Spouse Abuse Risk Assessment} (SARA)~\citep{kropp2005spouse} developed for the Canadian context \new{to design its domestic violence tool: VioGen}~\citep{pueyo2008valoracion, lopez2016eficacia}. In \new{fact, the Spanish tool is nowadays assessing gender based violence cases and has been reported to missclassify cases by women assessed by this algorithm.}\footnote{See: \textit{Victims denounce failures in VioGen, the algorithm against gender-based violence}. Available at \url{https://www.eldiario.es/tecnologia/victimas-denuncian-fallos-viogen-algoritmo-violencia-genero_1_8815201.html} (Last accessed April 8, 2024.).} \new{Despite these pitfalls,} VioGen is the most well-known tool for assessing the risk of intimate partner violence in Spain and \new{has been evaluated several times to improve its performance and calibrate the algorithm}. \new{Yet other risk assessment tools have drawn less attention, so they have not been scrutinised yet. This is the case of the EPV~\citep{echeburua2009assessing, echeburua2010escala}, the tool that is going to be analysed in this paper.} 

\section{\new{How is an algorithm introduced in the courtrooms to assess intimate partner violence? The case of the EPV tool in the Basque country}}
\label{sec:howepv}

\new{In 2022, 4,507 women officially reported their partners for violence in the Basque country, marking the highest number since 2009.\footnote{See: Report on violence against women, Basque Country. November 25, 2023. Available at: \url{https://en.eustat.eus/igualdad/nov252023/violencia.html} (Last accessed April 9, 2024).} These cases are evaluated automatically by a risk assessment tool: the EPV. This tool has been used by the Basque police\footnote{This law enforcement body is also referred as the Ertzaintza.} since 2009. Due to the lack of police officers that could evaluate the risk of femicide and severe violence by intimate partners, the law enforcement body decided to address the lack of human resources and ease their workloads by introducing this risk assessment tool. To do so, the police collaborated with  clinical psychology scholars based in the Basque country to develop the tool. The EPV was created from a sample of 1,081 perpetrators of partner violence who had been reported in a police station in the Basque country.}

\new{Building on a tool proposed to evaluate the risk of femicide in the Canadian context~\citep{kropp2005spouse}, the EPV was designed to analyse multiple risk factors, identifying individuals who had attempted to murder their partner or had committed severe violence against them, and another group of males who had committed less severe partner violence. Features with the most significant differences between the severe violence group (269 cases) and the less severe violence group (812 cases) in socio-demographic features were studied. These features were grouped into five psychometric categories: (1) personal data about the victim and the aggressor, (2) data about the relationship, (3) type of violence, (4) profile of the perpetrator, and (5) vulnerability of the victim (see Appendix \ref{app:a}). In this way, those risk factors that clearly discriminated between the two groups could be selected, and those with the highest capacity to discern for severe partner violence could be identified. The rest of the design, such as the number of \new{features} and the measurement scale, etc., are conditioned by the selection of items with the highest differentiation capacity and by the evaluation of each \new{feature} when it is present. If a \new{feature}  is not present, it would be scored as 0. If it is present, it would be scored as 1, 2, or 3 depending on the lower or higher predictive capacity that \new{feature} theoretically has. In the publication where the EPV was proposed, it is indicated that a 58-\new{features} interview was applied, selected based on the authors' clinical experience, literature review of previous studies, and contributions from the Basque police.} 

\new{The 1,081 perpetrators were interviewed by police officers at the time their partners filed the complaint~\citep{echeburua2010escala}. The police assigned the perpetrators to one of the two groups, interviewing both the perpetrators and the partners and taking into account the crime scene. Once all the questionnaires were completed, comparative analyses were conducted between the two groups to calculate the capacity of each \new{feature} to differentiate between severe and non-severe aggressors. This resulted in the 20 \new{features} that make up the questionnaire. The final \new{features} are those that showed the greatest capacity to differentiate between the two groups~\citep{echeburua2009assessing}.}

\begin{figure}
    \centering
    \includegraphics[trim={0 0 0 2cm},clip, width=0.8\textwidth]{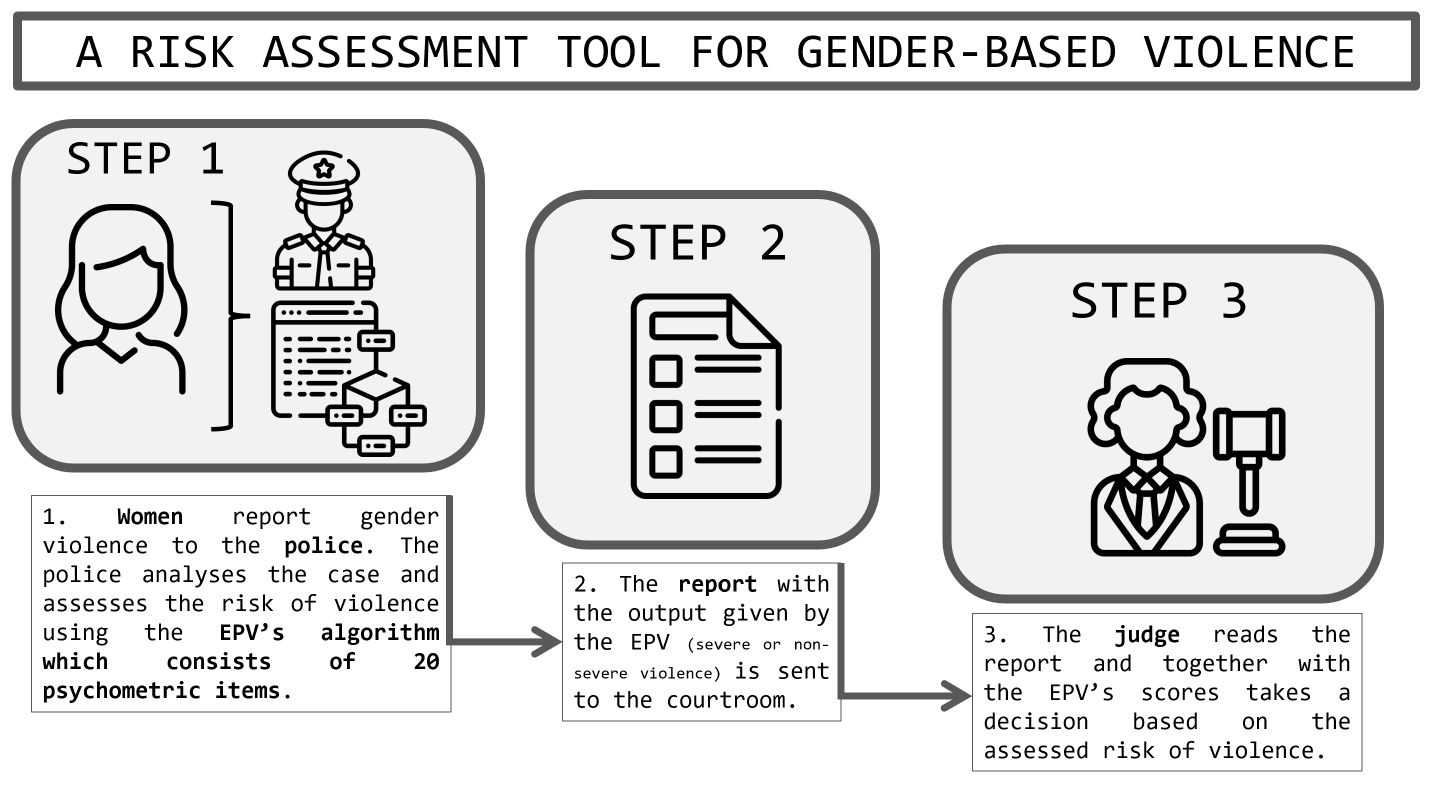}
    \caption{This diagram shows how the EPV, the risk assessment tool for intimate partner violence, is implemented in the Basque Country. First, the police assesses the case by evaluating 20 features that form the tool (for more details of the tool, see Appendix~\ref{app:b}) (\textbf{Step 1}). Second, with the EPV output together with a report of the case, the police sends the document to a courtroom (\textbf{Step 2}). Third, the judge evaluates and makes a decision based on the risk of severe violence oriented by the police report, which contains the EPV score, among others. In some cases, this report is accompanied by the victim's statement, as well as the perpetrator's, if they decide to testify. Additionally, some investigating judges must make the decision on what measures to take within an hour (\textbf{Step 3}).}
   \label{fig:epvr_method}
\end{figure}

\new{The developers of the EPV aimed at `provid[ing] non-clinical professionals with the prediction that allows the adoption of protection measures for victims, after being reported to police authorities'~\cite[p.~2]{echeburua2010escala} and `proposing a brief, easy-to-use scale [tool] that is practical for use by the police, social workers, forensic psychologists, and judges in their decision-making process'~\citep[p. 1055]{echeburua2008hay}. After its publication, the EPV was officially implemented in the law enforcement context of the Basque country following three main steps (see Figure~\ref{fig:epvr_method}): (1) a subject who reports domestic violence is interviewed by police authorities following a 20-\new{features} questionnaire \newnew{and estimates the risk of violence obtained by the EPV score} (see Appendix \ref{app:a}); (2) \newnew{the police sends the case report together with the algorithmic output the the courtroom and} (3) \newnew{the judge evaluates the risk of domestic violence given the report written by the police and the algorithmic output}. In 2010, the authors of the EPV proposed an updated version of the tool, the EPV-R~\citep{echeburua2010escala}. Rather than adding new data or assessing the risk factors, the authors suggested changing the rating score system from a binary scale (0-1) to a three-score scale (0-1-2)~\citep{echeburua2010escala}.}
 

\section{\new{The statistical and technical details of the EPV: Assessing its performance from a statistical perspective}}
\label{sec:tech}

\new{As illustrated in previous Section~\ref{sec:howepv},} the authors of the EPV trained this tool using 58 features of 1,081 cases. \new{These cases were reported between October 2005 to August 2006 in police stations~\citep{echeburua2009assessing}}. To build the risk assessment tool, the cases were labelled into severe (269 cases, 24.88\%) and non-severe violence (812 cases, 75.12\%) by the police. The \new{team of clinical psychology scholars} scrutinise the 58 items to find the 20 most relevant features to better distinguish between severe and non-severe cases implementing statistical comparative methods (see Appendix~\ref{app:a} for a description of the 20 \new{features} that composed the EPV). This comparative analysis relied on two well-known statistical hypothesis tests: the Student's $t$-test ($t$)\footnote{\new{The \textbf{Student's $t$-test ($t$)} is a statistical tool that determines whether the means of quantitative features of two groups are equal to each other. In the case of the EPV, the authors implemented this tool to compare differences on the mean age of aggressors between severe and non-severe cases~\citep{echeburua2009assessing}.}} and chi-squared test ($\chi^2$)\footnote{\new{The \textbf{Chi-squared test ($\chi^2$)} is a statistical tool that determines whether the means of qualitative features of two groups are equal to each other. In the case of the EPV, the authors implemented this tool to evaluate differences in age groups, nationality, profession, cultural or socio-economic level of aggressors between severe and non-severe cases~\citep{echeburua2009assessing}.}}. The $t$-test was applied to analyse statistically significant differences in the mean of scores of quantitative \new{features}. On the other hand, the $\chi^2$-test was applied to find significant differences on \new{qualitative} \new{features}. The reliability of the tool was assessed by calculating the Cronbach's alpha\footnote{\new{The \textbf{Cronbach's alpha}  is a statistical measure that assesses the degree of interrelatedness among a set of \new{features}. For example, in the context of the EPV, it was used to determine whether the 20 \new{features} were internally consistent in distinguishing between severe and non-severe aggressors.}} which analyses the internal consistency of the 20 \new{features} that are statistically useful to discriminate between severe and non-severe cases.

During the design process, \new{a profile comparison was implemented based on the demographics of male perpetrators of domestic} violence~\citep{echeburua2009assessing}. Analysing the profile of male aggressors, the statistical \new{test} results did not find any significant differences \new{among} age, profession, cultural level or socio-economic levels (see Table 2 in ~\cite[p. 931]{echeburua2008hay}). However, statistically significant differences were found in nationality: `foreign immigrant perpetrators, especially Latin Americans and Africans, committed more frequently (35.7\%) severe offenses than non-severe offenses (25.9\%)'~\citep[p. 930]{echeburua2009assessing}. In contrast, national statistics published by the Spanish government reveal that in 2008, Spanish males aggressors who commit femicides represented 63.2\%, whilst non-Spanish nationals represented 36.8\%.\footnote{Source: \url{https://violenciagenero.igualdad.gob.es/violenciaEnCifras/victimasMortales/fichaMujeres/home.htm} (Last accessed, January 20, 2022)\newnew{.}} \footnote{\new{From a statistical perspective, to further investigate how the EPV potentially perpetuate police bias, refer to: \textit{Biases with foreigners in the Ertzaintza's gender violence algorithm}. Available at: \url{https://www.elsaltodiario.com/policia/sesgos-extranjeros-del-algoritmo-violencia-genero-ertzaintza} (Last accessed April 10, 2024).}} \new{From a statistical perspective}, there is no evidence in the publication that this data could meet the usual requirements to implement Student's $t$-test.\footnote{Student's $t$-test should be applied under several \new{statistical} assumptions \new{on data}, such as the \new{data where the test is applied should follow a normal distribution}.} 

\new{After assessing differences in aggressors profiles, the authors implemented the same statistical tests, namely the Student's $t$-test and Chi-squared test ($\chi^2$), to assess mean disparities among the 20 \new{features}  between severe and non-severe cases. By analysing the rate of affirmative responses for each feature, the tests yielded statistically significant results for all 20 \new{features}, indicating differences among the 269 severe and 812 non-severe cases analysed. Surprisingly, nationality did not emerge as one of the most significant factors from these tests.\footnote{\new{The statistical test did show significant differences on the profile of aggressors given nationality.}} Factors such as intentional injuries, weapons, danger of death, and jealousy exhibited higher discrimination capacity (~\cite[Table 3, p. 932]{echeburua2008hay}).}


\new{The performance of risk assessment tools or machine learning classifiers can be assessed using various metrics. These metrics are computed based on the Confusion Matrix, which compares the number of correctly classified instances with those that are misclassified by the tool or algorithm. This matrix is composed of four main components: True Positives, True Negatives, False Positives, and False Negatives. True Positives (TP) and True Negatives (TN) indicate the correctly classified instances, while False Positives (FP) and False Negatives (FN) represent the incorrectly classified instances. These elements are used to calculate measures for evaluating the tool's efficiency. For instance, Accuracy is determined by summing the number of TP and TN across all instances. Sensitivity or True Positive Rate (TPR) calculates how many positive instances are correctly classified. On the other hand Specificity or True Negative Rate (TNR) calculates how many negative instances are correctly classified. In the case of the EPV, positive instances are severe cases and negative instances are non-severe cases. Its performance was assessed by evaluating the Accuracy of the tool setting different cutoff scores.\footnote{\new{The EPV is a risk assessment tool that assesses 20 items and outputs a final score between 0 and 20 that reflects the risk of severe violence, 0 meaning low risk and 20 highest risk. See Appendix~\ref{app:b}}.} Moreover, the trade-off between Sensitivity (or TPR) and Specificity (or TNR) was also examined. }

\begin{figure}
    \centering
    \includegraphics[width=0.8\textwidth]{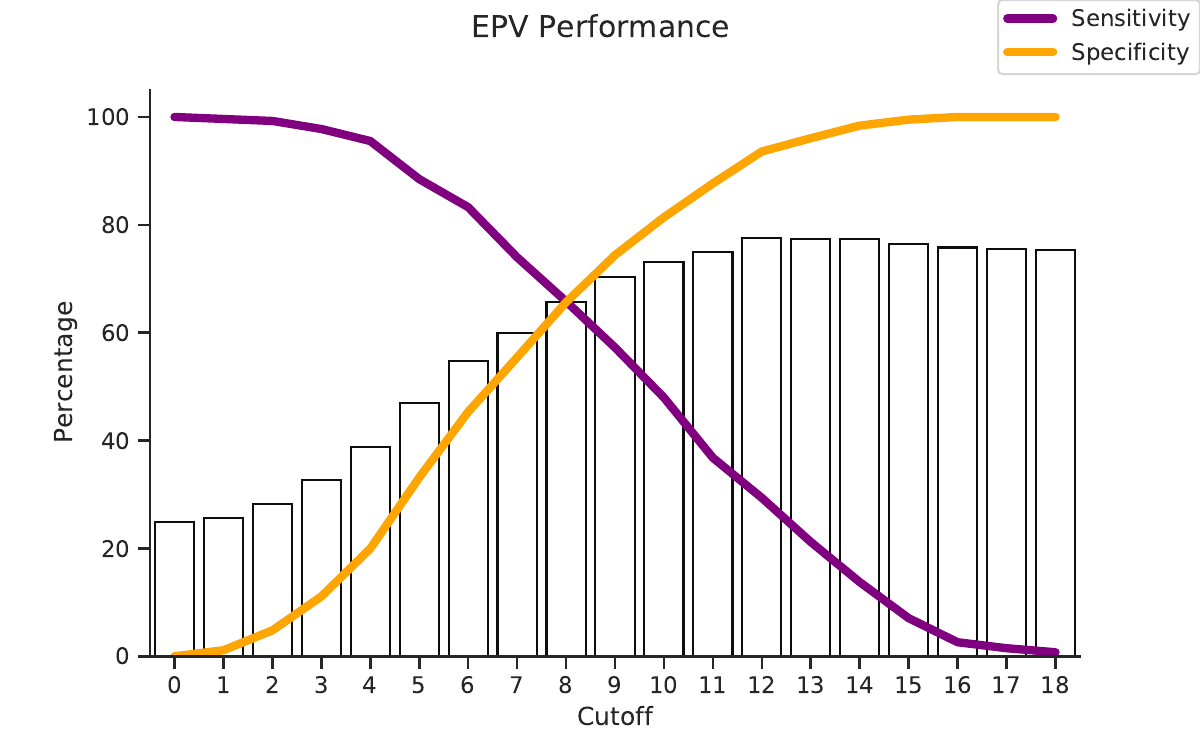}
    \caption{\new{Percentage of Accuracy (bars), Sensitivity (purple curve) and Specificity (orange curve) for every cutoff score on the EPV. At cutoff of 0 all cases are classified as severe; at cutoff of 18 all cases are classified as non-severe. This analysis was conducted to evaluate the most accurate cutoff score of the tool based on the trade-off between Sensitivity and Specificity, which was set at 10. At this cutoff, the EPV achieves 73.1\% of Accuracy, 47.96\% of Sensitivity (TPR) and 81.40\% Specificity (TNR). Given that Sensitivity is translated as the rate of correctly classified severe aggressors, 52.04\% are wrongly classified as non-severe aggressors.}  Source: Table 4 in \cite{echeburua2009assessing}.}
   \label{fig:curves}
\end{figure}

\new{Building on the results published by~\cite{echeburua2009assessing} (see Table 4 in p. 933), Figure~\ref{fig:curves} shows the percentage of Accuracy (bars), Sensitivity (purple curve) and Specificity (orange curve) for every cutoff score on the EPV. The cutoff indicates that cases evaluated by the EPV, scoring below the cutoff, are classified as non-severe cases, while those scoring above are classified as severe. For instance, at the cutoff of 0---cases below 0 are classified as non-severe and cases above 0 are classified as severe---all the cases are classified as severe. Then, the accuracy of the EPV is 24.9\% because only severe cases are correctly classified. This is translated into 100\% Sensitivity (TPR) (all positives, i.e. severe cases, are correctly classified) and 0\% Specificity (TNR) (all negatives, i.e. non-severe cases, are incorrectly classified). At the cutoff of 8, the EPV achieves 65.7\% Accuracy, which means that 710 cases are correctly classified by the tool (65.7\% of the 1,081 cases). At this cutoff, Sensitivity and Specificity curves intersect, achieving a percentage of 65\% in both cases. This means that at this cutoff, 65\% of severe and non-severe cases are correctly classified. At the cutoff of 18---cases below 18 are classified as non-severe and cases above 18 are classified as severe--- the EPV achieves 75.3\% Accuracy. However, the percentage of Sensitivity or TPR is almost 0---all severe cases are wrongly classified---and the percentage of Specificity or TNR achieves 100\%---all non-severe cases are correctly classified. In other words, as the authors of the EPV described:} 

\begin{quote}
    Thus, for example, a total score of 10, considered high risk, includes 48\% of the severe aggressors, which means that one half obtain lower scores, and only 18\% of the less severe aggressors obtain this score (false positive). If a stricter cutoff score had been chosen (e.g., 12), this would comprise 29\% of the severe cases, and there would be a much lower number of false positive (6\%), but at the cost of leaving out many severe aggressors (71\%; false negatives. In contrast, if a lower cutoff score had been chose (8 or 9), it would include not only a higher number of severe aggressors but also a large number of non-severe cases (false positives), which would limit the predictive capacity of the instrument.~\citep[p.~932]{echeburua2009assessing}
\end{quote}

\new{This analysis on Figure~\ref{fig:curves} set up the most efficient cutoff score for the tool, i.e. the score where police officers classify the case as severe or non-severe, which was proposed at 10. As claimed by the authors:}

\begin{quote}
    The proposed cutoff scores represent a reasonable equilibrium between the need to adequately detect the severe aggressors and the suitability of not extending this label to an unnecessarily high number of men who have behave violently against their partner, and those who, even though they committed an offence present a moderate or a low risk of carrying out severe behaviours that can place their partner's life at risk.~\citep[p.~933]{echeburua2009assessing}.
\label{Eche:quote}
\end{quote}

\new{However, there are other measures that also calculate errors and which the authors did not analyse~\citep{echeburua2009assessing}. For instance, the False Positive Rate (FPR) is the proportion of non-severe aggressors that are mistakenly classified as severe. On the other hand, the False Negative Rate (FNR) is the proportion of severe aggressors that are wrongly classified as non-severe.} In the context of domestic violence, FPR and FNR do not play the same role. \new{From a statistical perspective, it is desirable to obtain lower rates of FNR at the cost of higher rates of FPR. A FN in the EPV tool means that a severe case---where the aggressor can murder their partner---is classified as non-severe. Underestimating the risk of violence of an aggressor could imply putting women's life and that of their relative at serious risk.}  which implies that `reasonable equilibrium' between both error rates might not be desirable. \new{However, overestimating the risk of violence, i.e. classifying a non-severe case as severe (FP) could have also undesirable consequences for the aggressor.}

\new{At cutoff of 10, cases with 10 or fewer points in the EPV questionnaire (See Appendix~\ref{app:b}) are classified as non-severe, and as severe otherwise. With this classification, we obtain the confusion matrix of Table.~\ref{tab:conf_matrix}. This table shows that 129 severe aggressors were correctly classified, with a true positive rate (TPR) of 47.96\%, while 140 were incorrectly classified as non-severe, resulting in a false negative rate (FNR) of 52.04\%. In other words, 52.04\% of severe cases where the aggressor could potentially murder the partner are classified as non-severe. This means that a woman who has suffered severe violence and obtains 10 points on the EPV questionnaire, is less than half as likely to have her case will be assessed as high risk. In our view, this represents a high risk that could put women in danger as there the EPV shows a high rate of misclassifications for a tool that is currently used to evaluate intimate partner violence in the Basque Country.}

\new{Upon analysing the tool, we made another critical observation that the authors of the EPV did not explore: the significant imbalance between classes. In machine learning, the ratio of positive and negative classes directly affects the performance of a binary classification algorithm or a risk assessment tool, such as in this case. This appears to be the case with the EPV due to the high level of imbalance in the dataset, with 24.88\% severe and 75.12\% non-severe cases. In highly imbalanced data, commonly used metrics like Accuracy can produce misleadingly high performances by systematically predicting the majority class. This means that if there are significant classification errors in the minority class, they may not be adequately reflected in the accuracy percentage. For instance, the EPV's accuracy at 10 is reported as 73.1\% (see Figure~\ref{fig:curves}), yet there is a substantial disparity between the Accuracy in the minority (severe cases) and majority (non-severe cases), i.e. TPR and TNR: 47.96\% and 81.40\%, respectively.}



\begin{table}
\begin{tabular}{@{}cc cc@{}}
\multicolumn{1}{c}{} &\multicolumn{1}{c}{} &\multicolumn{2}{c}{\textbf{Predicted}} \\ 
\multicolumn{1}{c}{} & 
\multicolumn{1}{c}{} & 
\multicolumn{1}{c}{Predicted severe}& 
\multicolumn{1}{c}{Predicted non-severe} \\ 
\cline{2-4}
\multirow[c]{2}{*}{\rotatebox[origin=tr]{90}{\textbf{Actual}}}
&  Severe  & 129 (TP) & 140 (FN)   \\[1.5ex]
& Non-severe & 151 (FP)  & 661 (TN)  \\ 
\cline{2-4}
\end{tabular}
\vspace{0.5cm}
\caption{\new{EPV's Confusion Matrix at} cutoff score of 10. The number of FN (140) is higher than the number of TP (129), which implies that the tool is more likely to classify severe cases as non-severe when the obtained punctuation is 10. Source: \citep{echeburua2009assessing}.}
\label{tab:conf_matrix}
\end{table}

\section{Is judicial reasoning aided by EPV?}
\label{sec:legal}

The EPV appears to be, primarily, a tool for the police to assess the risk of dangerousness of a particular individual they encounter in domestic violence cases. In fast moving scenarios with limited information, police will use algorithmic tools to try to distribute their resources effectively and minimise risk, and EPV was developed by psychologists in an effort to assist. It is however frequently used in other situations. This paper considers the consequences of its use in the judicial process specifically, and what it might say for the use of similar risk-based algorithmic tools in court.

When deciding whether to prohibit the accused from approaching or communicating with an intimate contact, such as their wife or partner, judges will also have to make an assessment of risk. Indeed, where domestic violence is a potential factor, the assessment of objective risk is a legally-required element in judicial decisions restricting an individual’s movements, as set out below. 

A judge will hear representations of lawyers, documentary evidence, and sometimes oral evidence from witnesses, the accuser and the defendant.\footnote{For an exploration of where the output of an algorithms falls within this evidential framework see \cite{vanderstichele_normative_2019}.} 
The individual in the judicial role is also presented with \new{documentary} evidence of the EPV score, suggesting the risk level of the defendant. The score is presented without an algorithmic explanation. It is left to the judge to decide whether to follow this algorithmic evidence, and there is limited or no opportunity to interrogate the score at court. It may contrast strongly with their own judgement and human experience. How this conflict is resolved will vary according to the individual judge, but will have serious consequences for the parties to the case. We consider the status of the algorithm as a quasi-expert, and the tendency for humans to treat computer-derived scores with more weight.

\subsection{On what basis does a judge consider the assessment of risk?}

In the legal framework, and specifically Ley 544 ter 1 LECrim, judges are required to exercise their discretion in assessing the dangerousness of the offender:

\begin{quote} 

The Judges or Courts... taking into account the seriousness of the facts or the danger that the offender represents, may agree on their sentences... the imposition of one or more of the following prohibitions: 
a) The approach to the victim, or those of their relatives or other persons determined by the Judge or Court. 
b) That of communicating with the victim, or with those of his relatives or other persons determined by the Judge or Court. 
c) To return to the place where the crime was committed or to go to the place where the victim or his family resides, if they are different.\footnote{544 ter 1, \textit{Ley Enjuiciamiento Criminal} (LECrim).}

\end{quote}

 It is therefore relevant to consider to what extent the judge is likely to substitute their assessment of risk for that of the algorithm. European regulations highlight concerns with decisions made ‘solely on automated processing’; human decision-makers in these scenarios are required to exercise a real assessment of the merits.\footnote{European Parliament and Council, 2016:Art 22, https://gdpr-info.eu/art-22-gdpr/ (Last accessed, February 22, 2022).} The European Data Board warns that this cannot be circumvented by `fabricating human involvement' but these hybrid-type human interactions are where crucial focus needs to rest \citep{enarsson_approaching_2022}, a point reinforced recently by the Spanish Data Protection Board (Agencia Española de Protección de Datos).
 Considering the recent draft AI regulation proposed by the European Commission, even if we assume this kind of algorithm to be classified as high-risk, the draft regulation sees transparency, human oversight and accuracy as mitigating factors. This paper suggests that given the sphere being operated in, these terms may not provide sufficient protection; essential in this field is a nuanced and transdisciplinary approach. In any event, the path of progress for adoption of the framework within the draft regulation is uncertain, but it may finally be adopted in a modified form.\footnote{See for up-to-date progress of this regulation:   \url{https://www.consilium.europa.eu/en/press/press-releases/2022/12/06/artificial-intelligence-act-council-calls-for-promoting-safe-ai-that-respects-fundamental-rights/} (Last accessed, January 31, 2023).}

As we analyse this tool, it is also wise to have in mind the general principles common to many justice systems such as the rule of law, also codified in the Spanish Constitution. Article 1.1 asserts `Spain is hereby established as a social and democratic State, subject to the rule of law, which advocates as the highest values of its legal order, liberty, justice, equality and political pluralism'. Article 9 further asserts the principle of certainty that the rule of law will prevail.\footnote{The Spanish Constitution, 1978, \url{https://www.boe.es/legislacion/documentos/ConstitucionINGLES.pdf}. (Last accessed, February 22, 2022).} For various historical reasons, bound up with the status of the judicial court as final arbiter of fact and guilt, rather than being focused on generalised statistical risk, court decisions are individualised. They focus on determining the facts in the individual case, based on the evidence supplied\newnew{~\citep{oswald_technologies_2020}}.
The rule of law requires accessible, predictable judgements with clear justifications; on this the legitimacy of the system as a whole is built. 
 
\subsection{The Expert Algorithm?}

\begin{quote}
    ...something changed, something, I said to myself, \emph{this doesn't make sense}, there is some type of mistake here...\footnote{Direct and approved transcription of our discussion with a first-hand user, the judge in the Basque country.}
\end{quote}

How can the score of an algorithm be considered by a judge? 



Humans have been also used to evaluate the risks of violence but scholars view the expert with suspicion~\citep{GaraySuay2018}. The American Psychiatry Association affirmed the \emph{unreliability} of psychiatric predictions of longterm dangerousness as `now a recognized fact in the profession' in 1982 (APA,1982:5, as cited in~\citep[p.5]{GaraySuay2018}). `Experts' in the field might make assessments based on little data, but their authoritative tone, the `aura' of the expert, would often mean crucial questions were decided to the exclusion of judicial, or juror assessment. 

Given the a mount of data that is collected nowadays, algorithms have been imposed under the assumption that these tools are more reliable than humans. It is worthy of thorough exploration. It is not suggested in this paper that EPV will always be inferior to the judge's assessment, but justifying the use of algorithms assuming that a human will inevitably act as a fail-safe to an erroneous score is deeply problematic.\footnote{See also \newnew{work by} \citet{eubanks2018automating} on \textit{the Allegheny algorithm}.} The EPV's designers' openness is a positive example as it allows for far more scrutiny than in the case of some other algorithms. It illustrates how transdisciplinary approaches and user-engagement is key to highlighting risks and cautions regarding the implementation of a tool which was effectively designed for another purpose.

There is no question that, while originally designed for police use, the tool has been used in the courtroom, and such use has been acknowledged~\cite[p.~2]{echeburua2010escala}. There are \textit{prima facie} concerns with this usage, which are explored below. Further, to gain some insight into what this means in practice, we bring into this discussion an active judge in this jurisdiction with experience of EPV's use in the courtroom has provided their perspective, and agreed that translated excerpts our discussions be reproduced here. This single perspective, while not asserted to be representative of all instances, will hopefully be helpful to the reader. The above statement reflects his concerns when he evaluated an intimate partner violence. The EPV assessed the case a low risk but given his experience, he assessed the case as high risk. He did not understand why the algorithm was wrong. Nobody told him before how the algorithm was designed.


\subsection{Questioning the Expert}

In looking behind the algorithm, judges are faced with a number of obstacles.

\begin{quote}
    I can assess that there is sufficient risk to limit the freedom of movement, but the police say no, the risk is low. Of course, the lawyer can use this and say, look, the police say the risk is low. ... Now, I don't know what is behind this assessment, I don't know.\footnote{Discussion with the judge exposed to the EPV in considering the interrogation of the algorithm.}
\end{quote} 

The judge spoken to in the preparation of this paper confirmed that despite repeated attempts, they were unable to find out what the score had been based on, or where information had come from.\footnote{This judge discussed their frustrated attempts at trying to establish the basis of the assessment.} 
While the form designed to gather details considered relevant for the algorithmic assessment~\cite[p.74]{martinez2019}, judges may be thwarted in their examination of the specific circumstances and conduct of the accused. The bare fact of the EPV output is simply made known to users, likes judges, exposed to this tool.





\begin{figure}
    \centering
    \includegraphics[width=0.8\textwidth]{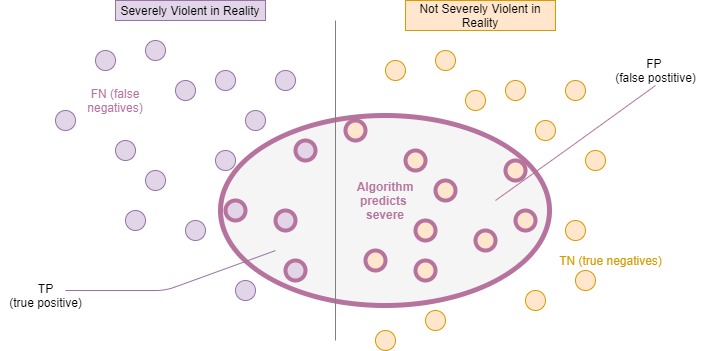}
    \caption{This diagram illustrates what it means to have a high number of FP and low number of TP predicted by an algorithm in this context. Those cases on the left side and \emph{outside} the ellipse are \emph{not} identified as high-risk (FN), which could imply that the number of severely violent individuals is underestimated. This will have immediate impact in an individual case where a violent aggressor is deemed low-risk.}
   \label{fig:drawTP_TN}
\end{figure}

 For instance, in the discussion held with the previous judge who observed the EPV's output in court reports, he reflected that he did not have experience of being referred to these papers in court, but if \new{indeed} a judge were referred to these sources they might be told that the algorithm is highly accurate, quoting statistical terms such as `specificity' or `sensitivity'~\citep{echeburua2009assessing}. These are not terms habitually used or instinctively understood in courtrooms. For current purposes, it may be helpful to understand that when EPV is described as accurate, this is in one particular calibration, where there is a relatively high level of false negatives, namely, a high number of individuals being identified as low risk despite being at high risk of inflicting severe violence \new{(see Figure~\ref{fig:drawTP_TN}, on page \pageref{fig:drawTP_TN} for an illustration of this)}. 
 \new{The reasoning regarding where the cut-off point was chosen and therefore where the line was draw, is discussed above (page \pageref{Eche:quote}). It is not the purpose of this paper to assert where exactly the line should be drawn in general terms. As Brian Christian highlights in his accessible account of ProPublica's critique of the COMPAS algorithm, it can in fact amount to `a conflict between two different mathematical definitions of fairness'~\citep[p. 69]{christian2021alignment}. What definition should be used, if at all, requires a transdisciplinary, contextual approach.}  Further, there is a dearth of studies that measure success of the algorithm as \textit{actual violence} as compared to projected risk. This means any assessment of accuracy is based on internal statistical assessments; real data is difficult to obtain, partially due to ethical constraints.\footnote{There is an ethical difficulty of randomised control trial as no action is not an option where risk is assessed as high. Without these objective measures of accuracy, the quality of the original information gathered by the police is even more highly relevant in the overall assessment of reliability.}
Given the specialised nature of these distinctions, a judge faced with what they consider to be a dangerous defendant, but a EPV rating of low risk, would be largely unaware of the true meaning of the algorithmic assessment, unless of course they have both considered the various academic papers on the subject and have experience of interpreting these kinds of statistics. Unlikely, given the pressured environment of the courtroom. 

\subsection{Behind the Expert} \newnew{T}here are a number of concerns raised when one analyses the inputs and mechanisms of the algorithm. For the courtroom judge, the enquiry cannot go beyond the `accuracy' of the tool itself. However, given the luxury of time and access to further information, this paper looks behind the algorithm. The appendix lists the 20 items that go into the EPV's overall assessment of risk (see Appendix~\ref{app:a}). Considering these in detail, the following can be said:  


\subsubsection{Human and subjective} Not all of the \new{features} are objectively measurable. In fact many of the elements are effectively subjective human judgements, being converted into input data despite the lack of clear parameters for choices. Item 6 is a composite measure, it asks the respondent to assess both seriousness and frequency \new{of behaviours}, meaning there could be inconsistent application, while Item 7 asks that a respondent assess if threats have been made and that the `profile' of the individual means they might perform those threats. \new{Far from being impersonal, these are highly subjective, personal assessments wrapped in an algorithmic cloak}. Items 4, 9, 10, 11 and 14 also seem vulnerable to this subjectivity (while Items 1 and 5 also appear problematic or insufficiently granular).
The problem with questions that allow for considerable discretion is that this allows for respondents to incorporate unconscious assumptions or prejudices. Of course judges are human too, and at risk of making assumptions, but behind the algorithm, all these uncertain aspects are masked.
 
In the sphere of judicial assessment of risk courtroom procedure seeks to mitigate this effect. It is of particular concern therefore that the items include uncertain concepts and definitions, which have counterparts in legal determinations. Item 9 asks the respondent to assess if previous conduct has shown `clear intention of causing severe or very severe injuries'. 
Essentially, we impute intent from actions, yet the tool does not prioritise identification of particular categories of actions in this item. Using `intent' as a measure risks respondents falling back on gut feeling, substituting a hard decision for an easy one, potentially based on their own unconscious assumptions~\citep{kahneman_by_2011}. 

\subsubsection{Absolute and Relative Risk}
Another cautionary factor is that the development of EPV, as described in the technical section, involved using a pool of individuals who had already been identified by reports. Therefore, it reflects relative risk: \textit{of those reported}, these are low-risk individuals, as they have not been predicted to commit severe violence. Low-risk within that pool of individuals is higher risk than if you had a pool of all individuals. This is something known by the designers of course, it is not hidden, but given the way in which the tool has been introduced, it is not clear to the judiciary, who could easily correlate low-risk with an absolute, almost negligible risk. Low-risk, given it is from a narrow pool of people who have been reported to police in some way and under suspicion of violent conduct is not analogous to innocent/non-violent, but crucially may be interpreted in this way by judges using the tool.


\subsubsection{Incomplete but determinative} Looking deeper into the algorithm not only are many of these elements filled out subjectively and inscrutably, but on occasion no answer is given at all. In an effort to improve EPV, the step was taken in EPV-R to give averaged scores in these cases \newnew{\citep{paniagua_EPVR_2017}. However, while \citeauthor{paniagua_EPVR_2017}} recognise the reasoning behind these issues and try to mitigate them (2017, p.~388), there is \textit{no indication to the judge} that crucial facts have been omitted, or the data is simply not present. Missing values in some areas may co-occur often (such as item 17 justification of violent behaviour and item 16 the presence of cruel behaviour and contempt~\citep{paniagua_EPVR_2017}). In the courtroom, the result is presented as equally reliable, if it is based on 20 items, or a much smaller number, and no indication is given that it represents an averaged rating. There is no indication of how limited the information was that went into the assessment of risk. As with the lie-detector test, these technologies are not necessary optimised for the purposes that a courtroom requires~\citep{oswald_technologies_2020}. 

\section{ \new{Statistical and legal limitations of the EPV}}
\label{sec:lim}

\new{The EPV has limitations that stem from how this system was statistically designed and legally implemented in the Basque court system.} The authors of the EPV publicly acknowledged that there are some limitations, \new{such as sample bias}~\citep{echeburua2010escala}. However, there is a lack of critical analysis on the risk/benefit balance of the EPV. Building on \new{an algorithmic accountability} framework and moved by the concern that this tool is being used today in \new{police stations and courtrooms in the Basque country}, we aim at highlighting its main \new{risks} from a \new{statistical} and legal perspective \new{and unveil how EPV is impacting on women's lives and courtrooms}.



\subsection{From a \new{statistical} perspective}

\subsubsection{False Negative Rate}
\new{The EPV's performance was assessed} with a focus on achieving a balance between True Positives (TP)---severe cases correctly classified as severe---and `True Negatives---non-severe cases correctly classified as non-severe---. Essentially, the authors of the EPV regarded the correct classification of severe and non-severe cases as equally significant.

From \new{an algorithmic} technical standpoint, the objective would be to maximise the correct classification of \new{positive or} severe cases, thereby minimising the False Negative Rate (TNR). This might result in a higher False Positive Rate  (FPR), i.e. non-severe cases being wrongly classified. However, given the low performance of the EPV, this would entail its own set of issues. Indeed, publicly available documents on the EPV~\citep{echeburua2009assessing} indicate a troublingly high False Negative Rates (FNR). For instance, in cases rated 10 points, the FNR for actual severe cases stands at 52.04\%. In other words, 5 out of 10 severe cases were inaccurately classified as non-severe. As the EPV currently assesses real cases of intimate partner violence, it's evident that this tool could significantly underestimating severe instances of domestic violence in the Basque Country. However, there is no data available to assess whether this limitation on the EPV could have entailed potential risks to partners.

\subsubsection{Efficiency}

\new{Efficiency} is a concept \new{employed to justify the deployment of algorithmic tools}. The notion of efficiency posits that (1) data-driven solutions surpass human capabilities and (2) can complete tasks more rapidly than humans. However, as ~\cite{stone1997policy}  argues: `efficiency does not tell you where to go [...]. Efficiency is a comparative idea'. Therefore, the efficiency of algorithms ought to be scrutinised: efficiency for whom and for what purpose? The EPV was introduced in pursuit of efficiency:


\begin{quote}
    The scale proposed to predict severe violence risk against a partner seems effective (with satisfactory psychometric properties) and efficient (short and easy to apply) for the goal sought [...]. This scale can be easily applied by personnel from the police, judicial, or social service settings, provided that they are sufficiently trained in its administration.~\cite[p.~934]{echeburua2009assessing}
\end{quote}

This tool was implemented in the pursuit of efficiency due to the shortage of human resources in police stations for assessing intimate partner violence. However, we must question the existing efficiency in light of the limitations identified. Moreover, algorithmic tools or risk assessment tools do not consistently outperform humans. Therefore, they should never be considered as a substitute for understaffing or lack of resources. For example, \cite{dressel2018accuracy} concluded that a commercial risk assessment software used in courtrooms to predict recidivism was no more accurate than individuals with no expertise in criminal justice. This study put into question the so-called efficiency of algorithms vs. humans. Whilst algorithms are faster than humans in identifying patterns and correlations within large datasets, the entire algorithmic lifecycle can prove to be a cumbersome and inefficient process. When an algorithm is deployed in a context as sensitive as gender violence, it is crucial to evaluate and audit it to ensure that its performance remains consistent. As the authors acknowledge in~\cite[p.~934]{echeburua2009assessing}: `the scale is only a photograph of a situation at a specific moment and should be completed with all the available data from the reality'. Therefore, the algorithm needs to be re-evaluated to address the statistical limitations.

\subsubsection{The making of severe and non-severe violence}

In the words of \citeauthor{hacking2007making}: `statistical analysis of classes of people is a fundamental engine' which has led to the `making up of people'~\citep[pp.~293-294]{hacking2007making}. Through the lens of numbers and quantification, intimate partner violence is now datafied, evaluated and scored to make up severe or non-severe cases. The quantification of social behaviour has been part of the long tradition of science research such as psychology or statistics~\citep{kropp2005spouse, gottfredson1988violence, miller1988predictions}. The numeralisation of the risk of intimate partner violence since the 1980s has transformed how academics, police authorities, judges, psychologists, psychiatrists and social workers assess its severity. 

In the case of the EPV, the risk of domestic violence is assessed based on 20 features purportedly capable of distinguishing between severe and non-severe cases. These 20 features were selected through statistical tests that revealed inherent differences across these groups. The authors of the EPV identified very intense jealousy' as a significant feature for distinguishing between cases. Indeed, 76.6\% of severe cases included in the tool's design exhibited very intense jealousy, according to the assessing police officer. Additionally, 55.9\% of non-severe cases also displayed intense jealousy. However, how did the police quantify `very intense jealousy'? Where is the qualitative and quantitative boundary between `very intense', `intense' and `non-intense' jealousy? How subjective is the assessment of jealousy? Given that individuals come from different backgrounds with varying political perspectives, quantifying items such as this one can lead to disparate evaluations. Two police authorities with different criteria regarding the concept of jealousy may assess an aggressor's controlling behaviour differently. Thus, quantifying behavioural items like this one can impact the design and performance of the tool and the statistical understanding of intimate partner violence.

\subsection{From a legal perspective}

\subsubsection{Two systems, no guidelines}

There is a lack of appropriate legal guidelines for the use of the EPV in a court scenario. There is no warning regarding the reliability of the data is given.\footnote{Discussions with the judge on the presentation of the algorithm's score.} The legal framework requires real deliberation by the judge, but the status of the algorithm as a quasi-expert closes down enquiry. Rules exist in legal proceedings to prevent reported speech being taken into account due to the disproportionate weight that can be attributed to hearing someone accusing another.\footnote{Heresay, known as \textit{testimonio indirecto} in Spanish law, is inadmissible unless substantiated by other evidence.} Equally, disproportionate weight can be attributed to these algorithmic utterances which contain within them a compound of uncertainties. Assumptions, trust in computer outputs, in a context of limited information and a pressured environment, call into question fundamental principles. Whether or not there is a warning or guideline for the judiciary (such as with the COMPAS algorithm in ~\citep{LoomisCase}), all parties are in the dark regarding how the assessment has been arrived at. If indeed acceptable for police use, this is particularly inappropriate for a hearing where different representatives will, of course, use the proclamation of the algorithm to argue their respective merits. 
Even a focus on transparency, human oversight and `accuracy'---all measures anticipated for high-risk algorithms under the draft EU regulation---will not necessarily be a panacea for this systemic issue.

\subsubsection{An unsatisfactory replacement}
In cases of partner violence, principles of the rule of law and due process require that individuals are aware of the case against them, and for the individual in the case, the right of free movement may be restricted. Equally, a victim's word may be doubted on the basis of a score of `low-risk'. We have highlighted that this assessment can be based on (1) incomplete information---gaps are filled in with averaged scores (2) subjective judgements with unclear parameters, allowing for assumptions and inconsistent application (3) an assessment of `risk' that is different to how a judge would conceptualise it. All this is given the impression of objectivity. The investigation here has shown that the judge is faced with a judgement with considerable impact for both victim and accused, and supplied with an algorithmic assessment which could be helpful, or could be actively misleading. As \cite{oswald_technologies_2020} says, this uncertainty as to the validity of algorithm's assessment is not consistent: two systems are colliding.

\subsubsection{Consequences of concern}

The consequences of a combination of factors in respect to EPV must give pause for thought regarding its comprehensive use in Basque courtrooms. Given `models are opinions embedded in mathematics' \citep{oneil_weapons_2016}, the assumptions made by those designing the tool will influence outcomes, even if all involved have the best intentions. The generalised, statistical approach has the potential to erode the individualised model of justice, which is a core element of the system.
More broadly, cooperation and trust in the system - the justice system and the algorithmic tool itself---will be eroded if participants feel they are not being treated with respect. People who see the system as illegitimate will not cooperate, will not come forward as witnesses / accusers, and those accused will be less likely to adhere to the rules of the system: \new{`[P]eople's reactions to legal authorities are based to a striking degree on their assessments of the fairness of the processes by which legal authorities make decisions and treat members of the public'~\citep[p. 284]{tyler_procedural_2003}.} Failure to grasp this context illustrates the need for an transdisciplinary approach, as has been pursued in this paper.

\section{\new{Conclusion}}
\label{sec:discussion}

Risk assessment tools have been largely criticised due to bias, disparate impact or opaque architecture in English speaking countries~\citep{angwin2016machine, eckhouse2019layers, singh2013authorship, stevenson2018algorithmic}. This paper advances this literature by critically scrutinising a tool implemented in the Basque courtrooms to evaluate intimate partner violence, moving the debate beyond simple issues of bias and opacity. Our analysis finds that the even non-sophisticated algorithms such as the EPV also entail risks and limits and could impact on domestic violence courtroom decisions. While some scholars have advocated for the use of algorithms in the court system, this work reveals that algorithmic failures could entail negative consequences~\citep{morin2023machine}.\newnew{~\citet{GaraySuay2018}} saw risk-assessing algorithms at court as something rare, treated with due caution by the judges involved;\footnote{The authors' greater concern in that paper was to inform policy decisions.} the experiences described by the member of the judiciary we spoke to suggest that in \new{2024} it is widespread and significant.\footnote{This judge reported it as taking place across courtrooms in the Basque country every day courts sat, as the algorithm is used by the police in each case, and these are common cases.} 

This tool goes to the very core of the judge's function on assessing domestic violence cases in the Basque country. If we do not accept that EPV is the best overall measure of risk under the legal framework, \new{judges} must weigh \new{their} assessment against their own judgement, based on the facts of the case. To do so, they must consider how reliable the EPV assessment is (as compared to the other evidence), and what harms or risks mean in this tool. 



This paper has been prepared by scholars from computer science and law to shed light on an algorithmic practice that has widely gone unreported, was not fully anticipated by the designers of the initial tool, and is so far unsupported by empirical research. In fact, we identify several elements that suggest the function of this algorithm in the decision-making of judges should be reconsidered, such as the high rate of false negative on severe cases or the dubious quantification of jealousy. Moreover, we emphasise the necessity to better understand the influence of risk assessment tools on judges~\citep{green2019disparate}. This work has not looked at the effectiveness for use by police, but given the above analysis, amalgamating these scores for use in court is highly problematic. The discussion with the judge who is exposed to this tool has further exposed how the EPV could also negatively impact on their decisions. Algorithmic hype and techno-solutionism could be at risk of being embraced by court systems, the EPV in its current implementation becoming a snake-oil tool~\citep{narayanan2019recognize} for judges to rely on. Moreover, \citeauthor{berber2023something} argued that the use of algorithms complicates the attribution of responsibility when something goes wrong: `It is an open question whether the advantages of the use of ML are worth the potentially irresolvable issues of moral responsibility in cases of harm caused by the decisions, and whether it is advisable to abandon the use of ML models for high-stakes decisions and go back either to human decision-making'~\citep[p.~11-12]{berber2023something}\newnew{.}

We have proposed an \new{algorithmic accountability} framework to analyse the EPV from a \new{statistical} and legal perspective. The experience of the judge who works in the Basque judicial system has contributed to better understanding the human-algorithm interaction.\footnote{Though discussions with \new{the judge exposed to this tool}, it is possible to point to some risks and concerns. To improve the roll out and assess the suitability of an algorithm such as EPV for judicial use, extensive discussions with judicial users and others would be more appropriate.} We propose an additional perspective that could potentially avoid negative consequences on the use of these tools and overcome risks, harms and limitations that have been identified through this paper. First, users (police agents, judges, prosecutors, social workers and healthcare professionals) of risk assessments tools should receive specific training on algorithms, so that they are able to identify algorithmic risks and harms, rather than relying too heavily on these tools~\citep{green2019disparate}. Second, we should reject any assumption that technology will automatically solve social and complex problems like. Risk assessment tools are implemented sometimes to address the lack of human resources. Yet algorithms and risk assessment tools cannot think like people, as\newnew{~\citet{dreyfus1989computers}} contended: `human beings have an intuitive intelligence that ``reasoning'' machines simply cannot match'. \new{Third, the Basque government should review this algorithm from a critical perspective to mitigate further harms~\citep{cobbe2021reviewable, verdegem2021ai}. There are ongoing academic debates within regarding algorithmic reparation to those negatively affected by such tools~\citep{rakova2023terms}.}
\new{Finally,} it is essential that judicial users understand their important overseeing role, that an algorithm---such as the EPV---cannot replace them and that its automatic assessment can be wrong. Failure to recognise this could lead to very serious consequences in cases of intimate partner violence\new{---such as María's case---}if they simply go by what the tool suggests.



\bibliographystyle{spbasic}      

%
%

\appendix
\section{Appendix: The 20 psychometric \new{features} of the EPV}
\label{app:a}

The following list presents a translation of the description of the \new{features}  found in the EPV's user guide to which the authors of this paper had access: 
\begin{enumerate}
 \item{ \textbf{Male aggressor or victim is an `immigrant'}}: Foreign origin is considered to exist when the aggressor or the victim originates from or is a national of a foreign country.
 
\item{ \textbf{Recently separated or in the process of separation}}: It is considered that there has been a recent separation or that separation proceedings are in progress when, in the last 6 months, the couple's relationship has undergone a crisis situation that provokes the beginning of the cessation of cohabitation, the beginning of separation proceedings or the existence of a judicial separation or divorce decision.
\item{ \textbf{Recent harassment of victim or breaking the restraining orders}}: In the last 6 months, bullying behaviour has taken place, which can manifest itself in the following forms: threatening phone calls, repeated forwarding of messages or continuous pressure on children.
\item{ \textbf{Existence of physical violence that can cause injuries}}: Any non-accidental conduct or act that causes or is likely to cause harm (pushing, hitting, hitting, burning, throwing objects, maiming, etc.). The means or instruments used in violent episodes are likely to cause injuries (knives, scissors, frying pans, irons, etc.).
\item{ \textbf{Physical violence in the presence of the children or other relatives}}: The aggressor has exposed his nature and does not care that his behaviour is known by the rest of the members of the family unit. He has overcome the inhibition of attacking in the presence of family members.
\item{ \textbf{Increase in the frequency and severity of the violent incidents in the past month}}: Violence is increasing and incidents (2 or more) are becoming more and more serious.
\item{ \textbf{Severe threats or threatening to kill in the past month}}: The threats are sufficient to make the victim feel frightened and submit to the aggressor's will. The profile of the aggressor suggests that he is likely to carry out her threats.
\item{ \textbf{Threatening with dangerous objects or with weapons of any kind}}: When threatened with any object or weapon likely to cause harm to the physical integrity of the person.
\item{ \textbf{Clear intention of causing severe or very severe injuries}}: The aggressor's attitude towards the victim, even if it does not materialise in serious injuries, denotes a clear intention to cause them, such as when an object is thrown against the victim's head, a sharp push is given, the victim is grabbed by the neck, thrown to the ground, etc.
\item{ \textbf{Sexual aggression in the couple relationship}}:
Any conduct or act of a sexual nature performed without the consent of the victim. The perpetrator uses intimidation methods (e.g. waking up children) to break the victim's will.
\item{ \textbf{Very intense jealousy or controlling behaviours toward partner}}: The aggressor feels very insecure in his relationship with the partner because he has an intense fear of losing his partner.
\item{ \textbf{History of violent behaviours with previous partner}}: The perpetrator has a history of physical or psychological violence with previous partners.
\item{ \textbf{History of violent behaviours with other people}}: The aggressor is (or has been) involved in violent incidents with other people in their family, social or work environment.
\item{ \textbf{Abuse of alcohol and/or drugs}}: The offender is abusing alcohol or drugs when he is currently using alcohol and/or drugs in a problematic way, either on a regular basis. In both cases, it is abusive use when it interferes negatively with the subject's behaviour towards the victim. However, this item does not assess positively in cases where there is habitual or sporadic use of drugs. Habitual or sporadic, but non-problematic, use below the limits of intoxication is not rated positively in this item, nor is intoxication or dependence without a clear effect on behaviour.
\item{ \textbf{History of mental illness and dropping out of psychiatric or psychological treatments}}: There is evidence that the perpetrator has a psychiatric history. There is evidence from information or direct evidence that he has abandoned treatment or that he has stopped taking the prescribed medication or therapy for treatment of his illness or that he has stopped taking the prescribed medication or therapy.
\item{ \textbf{Cruel, disparaging behaviors directed at the victim and lack of remorse}}: It is a style of behaviour of the aggressor that currently manifests itself in attitudes of contempt and humiliation, which leads the victim to feel subjugated, to which is added a lack of repentance. The aggression and violence of the person is exercised in an mechanical (non-emotional) and cold-blooded way, without being directly dependent on the situational circumstances (arguments, unpleasantness, etc.) that are behind the other type of violence.
\item{ \textbf{Justification of violent behaviour due to aggressor's own state or to victim's provocation}}: They use defence mechanisms when they offer their version of events: denial, justification, minimisation, etc. They blame the victim for causing them to be `forced' to use force. They do not consider themselves violent and perceive that they have been provoked by their partner.
\item{ \textbf{Victim's perception of danger of death in the past month}}: Assess the victim's perception when she has become aware that the perpetrator may kill (or seriously assault) her and feels in imminent danger of death (or serious assault). Probe on the basis of what facts the victim perceives this danger.
\item{ \textbf{Attempts to drop charges or going back on the decision to leave or report the aggressor to the police}}: The victim currently wishes the proceedings not to be initiated or closed for fear of reprisals. It is necessary to inquire about other motives that the victim may have that could cover up the fear of the aggressor: maintenance of the family unit, lack of economic resources, emotional dependence on her partner, shame in their social circle, etc.
\item{ \textbf{Victim's vulnerability because of illness, solitude, or dependence}}: The victim is alone and has no one (family or friends) to turn to in case of separation. Physical, economic or emotional dependency.
\end{enumerate}

\section{Appendix: The questionnaire of the EPV}
\label{app:b}

\begin{figure}
    \includegraphics[width=0.9\textwidth]{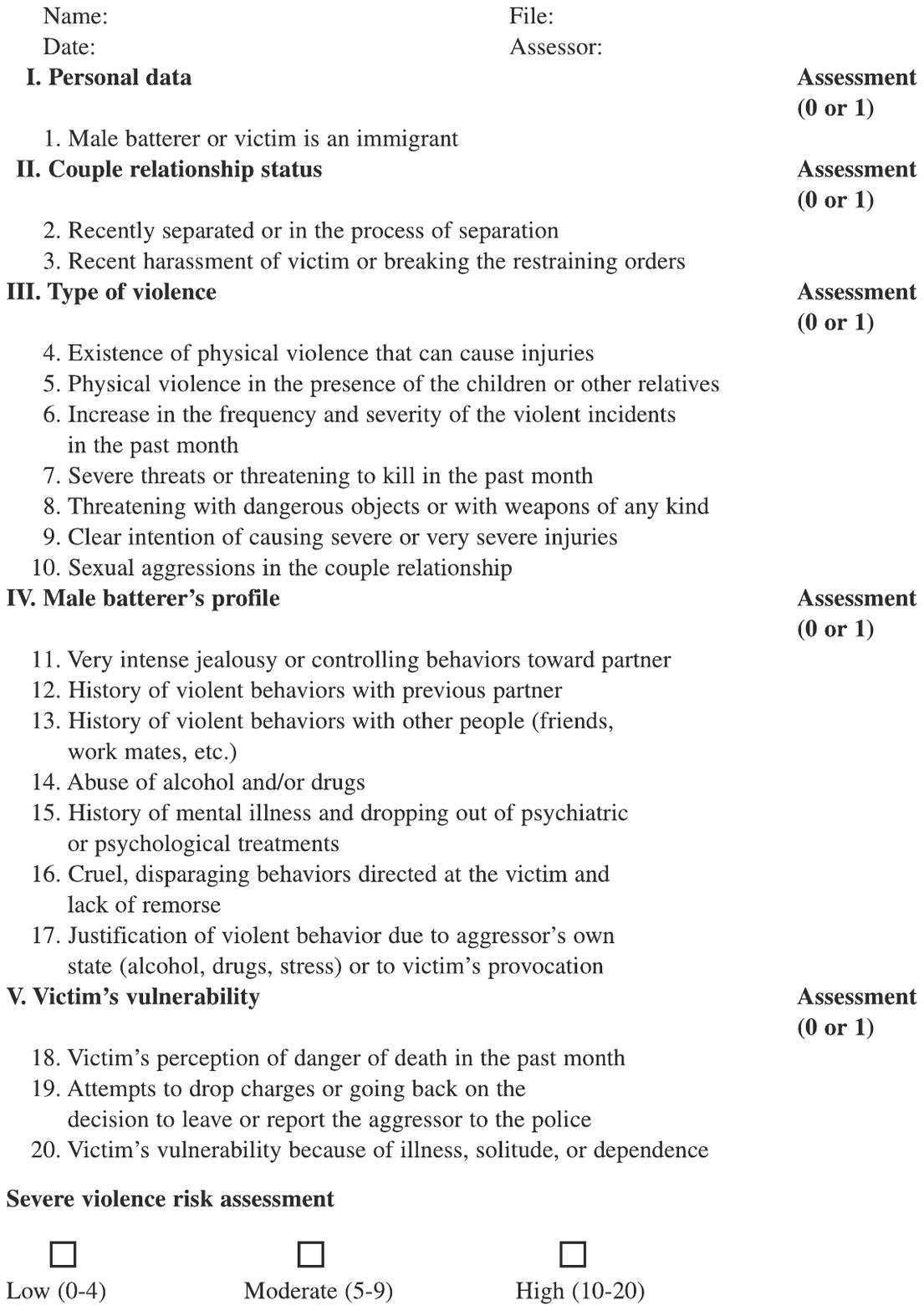}
    \caption{Questionnaire of the EPV to assess the risk of intimate partner violence in the Basque country (Spain). Source:~\cite{echeburua2009assessing}. \label{fig:epv_survey}}
\end{figure}

\end{document}